\documentclass[pre,aps,showpacs,nofootinbib,twocolumn]{revtex4}
\usepackage{amssymb}

\usepackage{graphicx}

\input{tcilatex}

\begin{document}

\author{$^{1}$M. L. Kuli\'{c} and $^{2}$O. V. Dolgov}
\title{ARPES kink is a "smoking gun" for the theory of high-T$_{c}$
superconductors:

dominance of the electron-phonon interaction with forward scattering peak}

\address{$^{1}$Johann Wolfgang Goethe-Universität Frankfurt am Main,
\\Theoretische Physik, Robert-Mayer-Str.8, 60054 Frankfurt/Main, Germany\\
$^{2}$Max-Planck-Institut f\"{u}r Festk\"{o}rperphysik,
Heisenbergstr.1, 70569 Stuttgart, Germany}

\begin{abstract}

The ARPES spectra in high-$T_{c}$ superconductors (HTSC) show
\textit{four distinctive features} in the quasiparticle
self-energy $\Sigma \mathbf{(k},\omega )$. All of them can be
explained consistently by the theory in which the electron phonon
interaction (EPI) with the forward scattering peak (FSP) dominates
over the Coulomb scattering. In particular, this theory explains
why there is \textit{no shift} of the nodal kink at $70$ $meV$ in
the superconducting state, contrary to the clear \textit{shift} of
the anti-nodal singularity at $40$ $meV$. The theory predicts a
\textit{``knee''-like} structure of $\mid Im\Sigma (\omega)\mid
=\mid Im\Sigma _{ph}(\omega)+Im\Sigma ^{C}(\omega)\mid$, which is
phonon dominated, $\mid Im \Sigma (\omega_{ph} )\mid \approx \mid
Im \Sigma_{ph} (\omega_{ph} )\mid \sim \pi
\lambda_{ph}\omega_{ph}/2$, for $\omega \approx
\omega^{(70)}_{ph}$, and shows linear behavior $\mid Im \Sigma
(\omega )\mid \approx \mid Im \Sigma_{ph} (\omega_{ph} )\mid +\pi
\lambda_{C,\varphi }\omega /2$ for $\omega
>  \omega^{(70)}_{ph}$ - due to the Coulomb scattering. ARPES
spectra give $\lambda_{ph}>1$ - which is obtained from $Re
\Sigma$, and $\lambda_{C}<0.4$ - obtained from $Im\Sigma$, i.e.
$\lambda_{ph} \gg \lambda_{C}$. The \textit{dip-hump} structure in
the spectral function  $A(\mathbf{k}_{F},\omega )$ comes out
naturally from the proposed theory.

\end{abstract}

\date{\today }
\maketitle

\textit{Introduction} - The pairing mechanism in high-temperature
superconductors (HTSC) is under intensive debate \cite{kulic}, \cite{allen2}.
In that respect ARPES experiments play a central role for
theory, since they give information on the quasiparticle spectrum,
life-time effects and indirectly the pairing potential. Recent
ARPES experiments on various HTSC families, such as $La_{2-x}Sr_{x}CuO_{4}$ and $BISCO$
\cite{lanzara}, \cite{damascelli}, \cite{zhou}, \cite{shen2}, show four distinctive features in
the quasiparticle self-energy $\Sigma \mathbf{(k},\omega )$:
\textbf{(I)} There is a \textit{kink} in the normal state quasiparticle spectrum, $\omega (\xi _{%
\mathbf{k}})$, in the \textit{nodal direction} $(0,0)-(\pi ,\pi )$ at the energy $%
\omega _{kink}^{(70)}\lesssim 70$ $meV$, which is a characteristic oxygen
vibration energy $\omega _{ph}^{(70)}$. However, the \textit{kink is not
shifted} in the superconducting state, contrary to the prediction of the
standard Eliashberg theory \cite{schrieffer}. The latter contains
integration over the whole Fermi surface and over the energy giving that
singularities in $\omega (\xi _{\mathbf{k}})$ (along all directions) must be
shifted in the superconducting state by the maximal gap value $\Delta _{0}$;
\textbf{(II)} In the \textit{anti-nodal region}, near $(\pi ,0)$ (or $(0,\pi
)$), there is a singularity in $\omega (\xi _{\mathbf{k}})$ in the normal
state at $\omega _{sing}^{(40)}\approx 40$ $meV$ - which is also a
characteristic oxygen vibration energy $\omega _{ph}^{(40)}$. This
singularity is \textit{shifted} in the superconducting state (at $T<<T_{c}$)
to $\omega \approx 60$ $meV(=\omega _{ph}^{(40)}+\Delta _{0})$, where $%
\Delta _{0}(\approx 20$ $meV)$ is the maximal superconducting gap
at the anti-nodal point. The experimental slopes of $Re \Sigma
\mathbf{(k},\omega )$  at the kink (and singularity) give the EPI
coupling constant $\lambda_{ph}>1$. The different shifts of
$\omega _{kink}^{(70)}$ and $\omega _{sing}^{(40)}$ in the
superconducting state we call the \textit{ARPES non-shift puzzle};
\textbf{(III)} There is a a \textit{``knee''-like} structure of
$\mid Im\Sigma (\omega)\mid =\mid Im\Sigma_{ph}(\omega)+ Im\Sigma
^{C}(\omega)\mid$, which is phonon dominated $\mid Im \Sigma
(\omega )\mid \approx \mid Im \Sigma_{ph} (\omega_{ph} )\mid \sim
\pi \lambda_{ph}\omega_{ph}/2$ with $\lambda_{ph}>1$ (obtained
from $Re\Sigma$) for $\omega \approx \omega^{(70)}_{ph}$, and for
$\omega > \omega^{(70)}_{ph}$ there is a pronounced linear
behavior of $\mid Im \Sigma (\omega )\mid \approx \mid Im
\Sigma_{ph} (\omega_{ph} )\mid +\pi \lambda _{C,\varphi }\omega
/2$ and $\lambda_{C}<0.4$ (obtained from $Im\Sigma$) - due to the
Coulomb scattering. It turns out that $\lambda_{ph}\gg
\lambda_{C}$; \textbf{(IV)} There is a \textit{dip-hump} structure
in the spectral function $A(\mathbf{k}_{F},\omega )$ with the
quasiparticle peak sharpening in the superconducting state near
the anti-nodal point.

Before explaining these distinctive features by the phonon-type
theory we stress, that the ARPES spectra especially the non-shift
puzzle and ``knee''-like structure, \textit{can not be explained
by the spin-fluctuation interaction} (SFI) due to the following
reasons: \textbf{(i)} the intensity of the SFI spectrum ($\sim
Im\chi (\bf Q,\omega )$ - the spin susceptibility at
\textbf{Q}=($\pi,\pi$)), although pronounced in slightly
underdoped materials, is strongly suppressed (even below the
experimental resolution) in the normal state of the optimally
doped HTSC oxides \cite
{bourges}, although their critical temperatures differ only slightly ($%
\delta T_{c}\sim 1$ $K$ !). Such a huge reconstruction of the SFI
spectrum around the optimal doping but with small effect on
$T_{c}$\ gives strong
evidence for the ineffectiveness of the SFI in pairing. \textbf{(ii)%
} The SFI theory \cite{norman} assumes unrealistically large coupling $%
g_{sf}\approx 0.65$ $eV$, (with the coupling constant $\lambda
_{sf}\approx 2.5$), while the ARPES \cite{lanzara},
\cite{damascelli}, \cite{shen2}, resistivity  \cite{kulic} and
magnetic \cite{kulkul} measurements give much smaller $g_{sf}\lesssim 0.1$ $eV$, i.e. $%
\lambda _{sf}<0.2<\lambda _{C}\lesssim 0.4$. Such a small $\lambda
_{sf}$ gives small $T_{c}$; \textbf{(iii)} if the kink at $70$
$meV$ in the normal state would be due to the magnetic spectrum,
then it would be strongly rearranged in the superconducting state,
contrary to the $ARPES$ results. On the other hand, the phonon
energies are only slightly ($\leq 5\%$) changed in the
superconducting state.; \textbf{(iv)} the magnetic resonance mode
at $41$ $meV$, which appears only in the superconducting state
\cite{norman}, can not cause the kink, since the latter is present
in $La_{2-x}Sr_{x}CuO_{4}$, where there is no magnetic resonance
mode at all \cite{lanzara}; \textbf{(v)} The non-shift puzzle and
``knee''-like structure of the ARPES spectra are related to phonon
features, which definitely disqualify the SFI and favor the
electron-phonon interaction (EPI) as the pairing mechanism. The
four distinctive features in the ARPES spectra can be explained by
the theory in which the electron-phonon interaction (EPI) with the
forward scattering peak (FSP) dominates over the Coulomb
interaction - the \textit{EPI-FSP model}.

\textit{The EPI-FSP model} - The central question for the EPI
theory is - why is the anti-nodal singularity $\omega
_{sing}^{(40)}$ shifted in the superconducting state, but the nodal kink $%
\omega _{kink}^{(70)}$ is not? We show that in order to solve the ARPES
non-shift puzzle one should go, as it is said before, beyond the standard Eliashberg theory for the
EPI \cite{schrieffer}. To remaind the reader, the standard
Eliashberg theory implies that $%
\omega _{sing}^{(40)}$ and $\omega _{kink}^{(70)}$ should be shifted in the
superconducting state to $\omega _{sing}^{(40)}\rightarrow \omega
_{ph}^{(40)}+\Delta _{0}$ and $\omega _{kink}^{(70)}\rightarrow \omega
_{ph}^{(70)}+\Delta _{0}$, respectively.
Here we show that the ARPES non-shift puzzle can be
explained by the EPI-FSP model which contains the following basic ingredients: (%
\textbf{1)} The \textit{EPI is dominant} in HTSC and
its spectral function $\alpha ^{2}F(\mathbf{k},\mathbf{k}^{\prime },\Omega )$%
, which enters the Eliashberg equations below, has a pronounced FSP
at $\mathbf{k}-\mathbf{k}^{\prime }=0$, due to
strong correlations. Its width is very narrow $\mid \mathbf{k}-\mathbf{k}%
^{\prime }\mid _{c}\ll k_{F}$ even for overdoped systems \cite{kulic2}, \cite
{kulic}. Near the Fermi surface one expects that $\alpha _{ph}^{2}F(\mathbf{k,k}%
^{\prime },\Omega )\approx \alpha _{ph}^{2}F(\varphi ,\varphi ^{\prime
},\Omega )$ \cite{allen}, and in strongly
correlated systems one has $\alpha _{ph}^{2}F(\varphi ,\varphi ^{\prime
},\Omega )\sim \gamma _{c}^{2}(\varphi -\varphi ^{\prime })$, where the
charge vertex $\gamma _{c}(\varphi -\varphi ^{\prime })$ is strongly peaked
at $\varphi -\varphi ^{\prime }=0$ with the width $\delta \varphi _{w}\ll
\pi $ \cite{kulic2}, \cite{kulic}. Thereby, one can put in leading order $\alpha
_{ph}^{2}F(\varphi ,\varphi ^{\prime },\Omega )\approx \alpha
_{ph}^{2}F(\varphi ,\Omega )\delta (\varphi -\varphi ^{\prime })$, which
picks up the main physics whenever $\delta \varphi _{w}\ll \pi $ \cite{kulic}%
. The EPI-FSP model, which results from the $t-J$ model with the electron-phonon
interaction \cite{kulic2}, \cite{kulic} predicts the following important results:
(a) the strength of
pairing is due to the EPI, while the residual
Coulomb interaction (including spin fluctuations) triggers the pairing to
d-wave one; (b) the transport coupling constant $\lambda _{tr}$ entering
the resistivity, $\varrho \sim \lambda _{tr}T$ is much smaller than the
pairing one $\lambda _{ph}$, i.e. $\lambda _{tr}<\lambda _{ph}/3$. We stress
that the FSP in the EPI of
strongly correlated systems is a general effect by \textit{affecting
electronic coupling to all phonons}. This is an important result, since for
some phonons (for instance the half-breathing modes of O ions)
the bare coupling constant $g_{0}^{2}(q)$ is peaked at large $%
q\sim 2k_{F}$ and therefore detrimental for d-wave pairing, while
the one renormalized by strong correlations
$g_{ren}^{2}(q)=g_{0}^{2}(q)\gamma _{c}^{2}(q)$ is peaked at much
smaller $q$, thus contributing constructively to d-wave pairing.
The Monte Carlo calculations on the Hubbard model with the EPI and
finite repulsion \cite{hanke}, confirm the existence of the FSP in
the EPI - previously found analytically in \cite{kulic2};
(\textbf{2}) the dynamical part (beyond the Hartree-Fock) of the
Coulomb interaction is characterized by the spectral function $S_{C}(\mathbf{%
k},\mathbf{k}^{\prime },\Omega )$. The ARPES non-shift puzzle implies that $%
S_{C}$ is either peaked at small transfer momenta $\mid \mathbf{k}-\mathbf{k}%
^{\prime }\mid \ll k_{F}$, or it is so small that the shift is weakly
affected and below the experimental resolution of ARPES. Since the ARPES
data give also that the Coulomb coupling constant $\lambda _{C}<0.4$ is much
smaller than $\lambda _{ph}>1$, then the kink is practically insensitive
to the $\mathbf{k}$-dependence of $S_{C}$. Due to simplicity we assume the former case -
see also discussion after Eq.(4); (\textbf{3}) The scattering potential due to
non-magnetic impurities has pronounced forward scattering peak - due to
strong correlations \cite{kulic2}, \cite{kulic}, thus making d-wave pairing
robust in the presence of impurities - see more below.

\textit{Eliashberg equations for the EPI-FSP model} - The Matsubara Green's
function is defined by ($k=(\mathbf{k},\omega _{n})$)
\begin{equation}
G_{k}=\frac{1}{i\omega _{k}-\xi _{\mathbf{k}}-\Sigma _{k}(\omega )}=-\frac{i%
\tilde{\omega}_{k}+\xi _{\mathbf{k}}}{\tilde{\omega}_{k}^{2}+\xi _{\mathbf{k}%
}^{2}+\tilde{\Delta}_{k}^{2}},  \label{1}
\end{equation}
where $\xi _{\mathbf{k}}$, $\tilde{\omega}_{k}$ and $\tilde{\Delta}_{k}$ are
the bare quasiparticle energy, renormalized frequency and gap, respectively
\cite{allen}. The 2D Fermi surface of HTSC is parameterized by $\mathbf{k}%
=(k_{F}+k_{\perp },k_{F}\varphi )$, where $k_{F}(\varphi )$ is the Fermi
momentum and $k_{F}\varphi $ is the tangent on the Fermi surface \cite{allen}%
. In that case $\xi _{\mathbf{k}}\approx v_{F,\varphi }k_{\perp }$ and $\int
d^{2}k[...]\approx \int \int d\xi d\varphi k_{F,\varphi }/v_{F}(\varphi
)=\int \int N_{\varphi ,\xi }d\xi d\varphi $. After the $\xi $-integration
the Eliashberg equations in the FSP model read

\begin{equation}
\tilde{\omega}_{n,\varphi }=\omega _{n}+\pi T\sum_{m}\frac{\lambda
_{1,\varphi}(n-m)\tilde{\omega}_{m,\varphi }}{\sqrt{\tilde{\omega}%
_{m,\varphi }^{2}+\tilde{\Delta}_{m,\varphi }^{2}}} +\Sigma^{C} _{n,\varphi},
\label{2}
\end{equation}

\begin{equation}
\tilde{\Delta}_{n,\varphi }=\pi T\sum_{m}\frac{\lambda _{2,\varphi }(n-m)%
\tilde{\Delta}_{m,\varphi }}{\sqrt{\tilde{\omega}_{m,\varphi }^{2}+\tilde{%
\Delta}_{m,\varphi }^{2}}}+\tilde{\Delta}_{n,\varphi }^{C},  \label{3}
\end{equation}
where $\lambda _{1(2),\varphi }(n-m)=\lambda _{ph,\varphi }(n-m)+\delta
_{mn}\gamma _{1(2),\varphi }$ with the electron-phonon coupling function $%
\lambda _{ph,\varphi }(n)$
\begin{equation}
\lambda _{ph,\varphi }(n)=2\int_{0}^{\infty }d\Omega \frac{\alpha
_{ph,\varphi }^{2}F_{\varphi }(\Omega )\Omega }{\Omega ^{2}+\omega _{n}^{2}}.
\label{4}
\end{equation}

Note, that Eqs.(2-3) have a \textit{local form} as a function of the angle $%
\varphi $, i.e. the energies at different points on the Fermi surface are
decoupled. Just this (decoupling) property of the Eliashberg equations
in the EPI-FSP model, is crucial for solving the ARPES
non-shift puzzle. The term $\Sigma _{n,\varphi }^{C}$ is due to the
dynamical Coulomb effects and its calculation is the most difficult part of
the problem.
$\Sigma^{C}$ is proportional to the charge vertex $\gamma _{c}(\varphi
-\varphi ^{\prime })$ and, as we said in (\textbf{2}), we assume that it
is also almost ''local'' on the Fermi surface, although this
assumption is not crucial at all, since $\lambda _{C}\ll \lambda _{ph}$.
After the $\xi $-integration it reaches the same form as the second term in $%
Eq.(2)$, where $\lambda _{1,\varphi }(n-m)$ is replaced by the Coulomb
coupling function $\lambda _{C,\varphi }(n-m)$. The latter has the same form
as Eq.(4) but $\alpha _{ph,\varphi }^{2}F_{\varphi }(\Omega )$ is replaced
by $S_{C,\varphi }(\Omega )$. ARPES spectra give evidence that $Im \Sigma
_{\varphi }^{C}(\omega )\approx -\pi \lambda _{C,\varphi }\omega /2$ at $%
T<\omega <\Omega _{C}$ which we reproduce by taking $S_{C,\varphi }(\omega
)=A_{C,\varphi }\Theta (\mid \omega \mid -T)\Theta (\Omega _{C}-\mid \omega
\mid )$, where $A_{C,\varphi }$ is normalized to obtain $\lambda _{C,\varphi
}\lesssim 0.4$.
The contribution of the Coulomb interaction to the gap, $\tilde{\Delta}_{n,\varphi }^{C}$,
in $Eq.(2)$ includes the following
effects: (i) of the Hartree-Fock pseudopotential - which maximizes $T_{c}$
when $<\tilde{\Delta}_{n,\varphi }>_{F}=0$ and favors unconventional (d-wave)
pairing; (ii) of the dynamical part of the Coulomb interaction which is
unknown and an approximation for $\tilde{\Delta}^{C}$ is needed. The SFI
approach assumes that $\tilde{\Delta}^{C}(\mathbf{k},\omega _{n})$ depends
on the dynamical spin susceptibility $\chi _{s}$. Since $Im\chi _{s}(\mathbf{%
q},\omega )$ is peaked at $\mathbf{Q}=(\pi ,\pi )$ this term is
repulsive and favors d-wave pairing. Although
$\tilde{\Delta}_{n,\varphi }^{C}$ contributes little to
$\tilde{\Delta}_{n,\varphi }$, it is important to trigger
superconductivity from s- wave to d-wave pairing \cite{kulic},
\cite{kulic2}.

In $Eqs.(1-2)$ non-magnetic impurities are included. Strong
correlations induce the FSP in the impurity
scattering matrix, being $t(\varphi ,\varphi \prime ,\omega )\sim \gamma
_{c}^{2}(\varphi -\varphi ^{\prime })$ \cite{kulic}. In leading order one has $%
t(\varphi ,\varphi \prime ,\omega )\sim \delta (\varphi -\varphi ^{\prime })$%
, thereby not affecting any pairing. In reality impurities are
pair-breaking for d-wave pairing and the next to leading term is
necessary. This term is controlled by two scattering rates,
$\gamma _{1,\varphi }$ and $\gamma _{2,\varphi }$, where $\gamma
_{1,\varphi }-\gamma _{2,\varphi } \geq 0$. The case
$\gamma_{1,\varphi }=\gamma _{2,\varphi }$ leads to the extreme
forward scattering - not affecting $T_{c}$, while $\gamma
_{2,\varphi }=0$ means an isotropic and strong pair-breaking
scattering \cite{kulic}.

\textit{Quasiparticle renormalization} - The quasiparticle energy
$\omega (\xi _{\mathbf{k}})$ is the pole of the retarded Green's
function. For numerical calculations we take for simplicity the
Lorentzian shape for $\alpha _{ph,\varphi }^{2}F_{\varphi }(\Omega
)$ centered at $\omega_{ph}$. Since our aim is a qualitative
explanation of the ARPES non-shift puzzle, we perform calculations
only for moderate coupling constants $\lambda _{ph,\varphi
}\approx \lambda _{ph}=1$, $\lambda _{C}=0.3$ in both, the nodal
and anti-nodal direction. In fact they can take larger values,
i.e. $\lambda _{ph}\lesssim 2$  especially in the anti-nodal
region. It is apparent from $Eqs.(1-2)$ that the quasiparticle
renormalization is local (angle-decoupled) on the Fermi surface.
This behavior is expected to be realized in a more realistic model
with the finite width $\delta \varphi _{w}$ when $\delta \varphi
_{w}\ll \pi $ \cite {kulic}.

\textbf{(I)} \textit{Kink in the spectrum in the nodal direction }
- The kink at $\omega _{kink}^{(70)}\approx 70$ $meV$ in $\omega (\xi _{\mathbf{k}%
})$ means that the quasiparticles moving along the nodal direction $(\varphi =\pi /4)$
interact with phonons with frequencies up to 70 meV \cite{carbotte}, i.e. $\alpha
_{ph,\pi /4}^{2}F_{\pi /4}(\Omega )\neq 0$ for $0<\Omega \lesssim 70$ $meV$.
Since $\Delta _{\pi /4}(\omega )=0$ then the ''local'' form of $Eq.(2)$
implies that the spectrum $\omega (\xi _{\mathbf{k}})$ is \textit{not shifted%
} in the superconducting state. Numerical calculations in $Fig.1a$ confirm
this analytical result what is in agreement with ARPES results \cite{lanzara}.
\begin{figure}[tbp]
\includegraphics*[width=8cm]{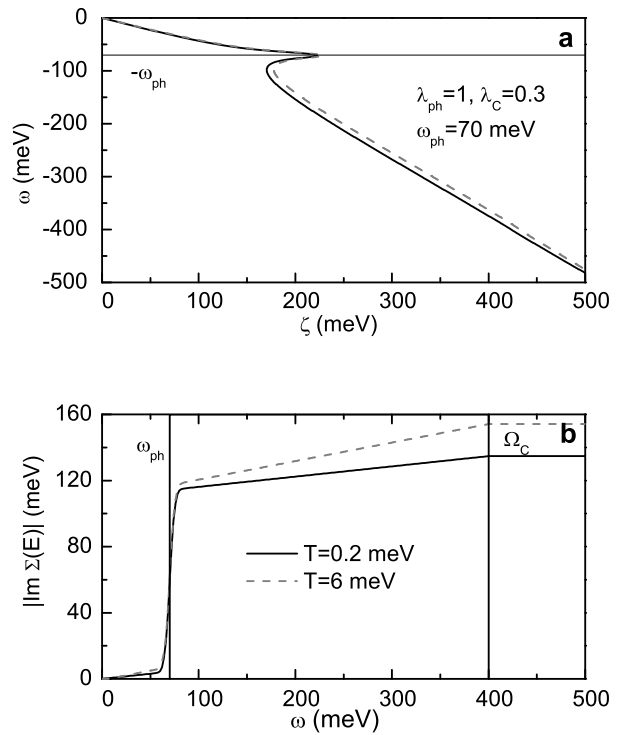}
\caption{\textbf{(a)} The quasiparticle-spectrum $\protect\omega (\protect\xi
_{\mathbf{k}})$ and \textbf{(b)} the imaginary self-energy $Im\Sigma (%
\protect\xi =0,\protect\omega )$ in the nodal direction ($\protect\varphi =%
\protect\pi /4$) in the superconducting ($T=0.2$ $meV$) and normal ($T=6$ $%
meV$) state. $\Omega _{C}=400$ $meV$ is the cutoff in $S_{C}$.}
\label{1}
\end{figure}
It is expected that for a realistic phonon spectrum the theoretical singularity in $\omega (\xi _{%
\mathbf{k}})$ (shown in $Fig.1a$) will be smeared having also an
additional structure due to other phonons which contribute to
$\alpha^{2}F(\omega)$.

\textbf{(II)} \textit{Singularity in the anti-nodal direction }
- The singularity (not the kink) in $\omega (\xi _{\mathbf{k}%
})$ at $\omega _{sing}^{(40)}$ in the anti-nodal direction
($\varphi \approx \pi /2)$ is observed in ARPES in the normal and
superconducting state of $La_{2-x}Sr_{x}CuO_{4}$ and $BISCO$
\cite{shen2}. This means that the quasiparticles moving in the
anti-nodal direction interact with a narrower phonon spectrum
centered around $\omega _{ph}^{(40)}\approx 40$ $meV$. Since $\mid
\Delta _{\pi /2}(\omega )\mid =\Delta _{0}$, then
$Eq.(1)$ gives that in the normal state $\omega (\xi _{\mathbf{k}%
})$ is singular at $\omega _{sing}=\pm \omega _{ph}^{(40)}$, while
in the superconducting state \textit{the singularity is shifted}
to $\omega _{sing}^{(40)}=\pm (\omega _{ph}^{(40)}+\Delta _{0})$.
This is confirmed by numerical calculations in $Fig.2a$ for $\omega (\xi _{%
\mathbf{k}})$, and in $Fig.2b$ for $Im\Sigma (\varphi ,\omega )$, for $%
\lambda _{ph}=1$ and $\lambda _{C}=0.3$. ARPES spectra give
$\lambda _{ph}>1$ in the anti-nodal region. \cite{shen2}.

Note, that the theoretical singularity in $Fig.1a$ is stronger
than in $Fig.2a$, because the calculations are performed for the
same temperature, and since $\omega _{ph}^{(70)}>\omega
_{ph}^{(40)}$ the latter singularity is smeared by temperature
effects more than the former. The real shape of these
singularities depends on microscopic details, such as for instance
the presence of the van Hove singularity slightly below the Fermi
surface in the anti-nodal region, etc. This will be studied
elsewhere.

\textbf{(III)} \textit{The ``knee''-like shape of $Im\Sigma (\xi
=0,\omega )$} - The ``knee'' is shown in Fig.1b for the nodal kink
(at $\omega_{ph}=70$ $meV$) and in Fig.2b for the antinodal
singularity (at $\omega_{ph}=40$ $meV$). In both cases there is a
clear ``knee''-like structure for $\omega $ near $\omega _{ph}$,
what is in accordance with the recent ARPES results in various
HTSC families \cite{lanzara}, \cite{damascelli}, \cite{zhou},
\cite{shen2}. From Fig.1b it is seen that
for energies $\omega _{ph}^{(70)}<\omega <\Omega _{C}$ the linear term is discernable in $%
\mid Im\Sigma \mid =\mid Im\Sigma _{ph}+ Im\Sigma ^{C}\mid \sim
\mid Im\Sigma _{ph}(\omega_{ph})\mid +\pi \lambda _{C,\varphi
}\omega /2$, while for $\omega \approx\omega _{ph}^{(70)}$ the
slope of $\mid Im\Sigma (\xi =0,\omega )\mid$ is steeper, since
for $\lambda _{ph}(=1)\gg \lambda _{C}(=0.3)$ the term $\mid
Im\Sigma _{ph}(\omega_{ph} )\mid (\gg \mid
Im\Sigma^{C}(\omega_{ph} )\mid )$ dominates. The ``knee''-like
shape of $Im\Sigma (\xi =0,\omega )$, as well as the non-shift of
the kink at 70 meV, are ''smoking gun'' results for HTSC theories,
which obviously favor the EPI as the pairing interaction. At
present only the EPI-FSP theory is able to explain all distinctive
features in ARPES spectra in a consistent way. The ``knee''-like
structure in the normal state was also obtained in \cite{greco},
where the EPI and Coulomb interaction are treated
phenomenologically. The EPI-FSP theory predicts also the
``knee''-like structure in the anti-nodal region. However, in this
case the closeness of the anti-nodal point to the van Hove
singularity may influence $\Sigma _{ph}(\xi =0,\omega )$
significantly and change its shape too. This will be studied
elsewhere.

\begin{figure}[tbp]
\includegraphics*[width=8cm]{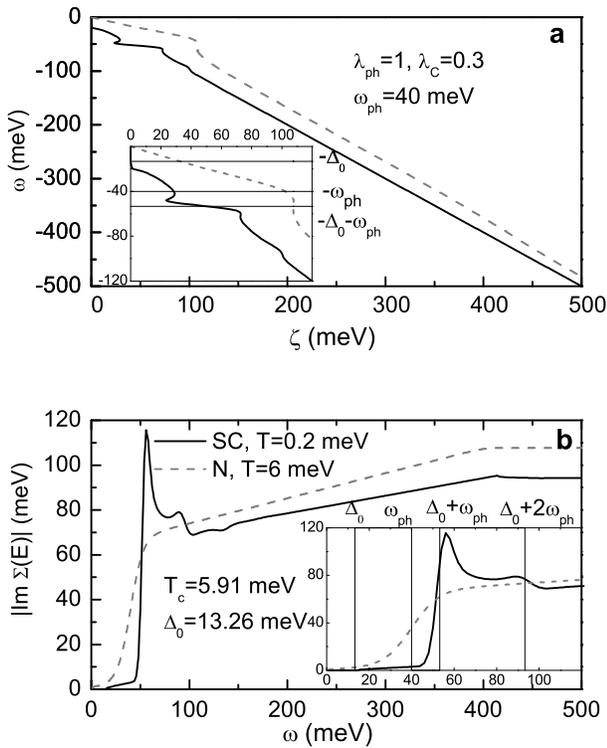}
\caption{\textbf{(a)} The quasiparticle-spectrum $\protect\omega (\protect\xi%
_{\mathbf{k}})$ and \textbf{(b)} the imaginary self-energy $Im \Sigma (%
\protect\xi =0,\protect\omega )$ in the anti-nodal direction ($\protect%
\varphi =0;\protect\pi /2$) in the superconducting ($T=0.2$ $meV$) and
normal ($T=6$ $meV$) state.}
\end{figure}

\textbf{(IV)} \textit{ARPES dip-hump structure } - The EPI-FSP model explains
qualitatively the dip-hump structure in $A(\varphi ,\omega )=-ImG(\varphi
,\omega )/{\pi }$ which was observed recently in ARPES \cite{damascelli}. In
$Fig.3a$ it is seen that the dip-hump structure is realized in the normal
state (also in the presence of impurities) already for a moderate coupling constant $%
\lambda _{ph}=1$. The dip is more pronounced in the superconducting state
where the peak in
$A(\omega )$ is appreciable narrowed, what is in accordance with ARPES
experiments \cite{damascelli}.
\begin{figure}[tbp]
\includegraphics*[width=8cm]{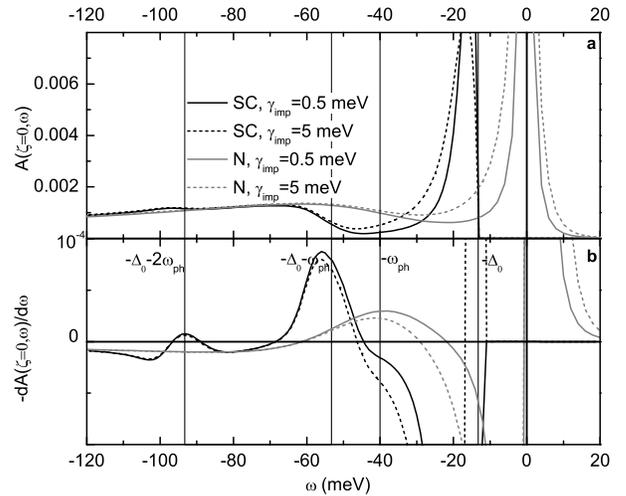}
\caption{\textbf{(a)} The spectral function $A(\protect\xi =0,\protect\omega
)$ and \textbf{(b)} - $-dA(\protect\xi =0,\protect\omega )/d\protect\omega $
in the anti-nodal direction in the superconducting ($T=0.2$ $meV$) and
normal ($T=6$ $meV$) state for various impurity scattering rate $\protect%
\gamma _{1}$ and $\protect\gamma _{2}=0$; $\protect\lambda _{ph}=1$, $%
\protect\lambda _{C}=0.3$.}
\end{figure}
Contrary to expectations, the dip-energy does not coincide with
the (shifted) phonon energy at $\omega _{ph}=40$ $meV$. However,
the positions of the maxima of $-dA/d\omega $ appear near the
energies ($-\Delta _{0}-n\omega _{ph}$) as it is seen in $Fig.3b$.
The calculations give also a dip in both, the anti-nodal and nodal
density of states $N(\omega )$ (not shown) already for $\lambda
_{ph}=1$, which is more pronounced for larger $\lambda _{ph}(>1)$.

\textit{Discussion and conclusions} - In obtaining $Eqs.(1-4)$ in
the EPI-FSP model the Migdal vertex corrections due to the electron-phonon
interaction are neglected. In \cite{pietro} it is shown that these corrections may
increase $T_{c}$ significantly, at the same time decreasing the isotope effect, even for $\lambda _{ph}<1$.
However, these results can not change the qualitative picture obtained by the
present theory.

In conclusion, the four distinctive features in the quasiparticle
self-energy $\Sigma \mathbf{(k},\omega )$, obtained from the ARPES
spectra in HTSC materials, are explained consistently by the
theory in which the electron phonon interaction (EPI) with the
forward scattering peak (FSP) dominates over the Coulomb
scattering. In particular, this theory explains why there is
\textit{no shift} of the nodal kink at $70$ $meV$ in the
superconducting state, contrary to the clear \textit{shift} of the
anti-nodal singularity at $40$ $meV$. The ``non-shift'' effect is
a direct consequence of the existence of the FSP in the EPI, i.e.
due to the long-range character of the electron-phonon interaction
in HTSC oxides \cite{kulic3}. This is also supported by the
pronounced long-range Madelung EPI, which is caused by the
ionic-metallic character of layered HTSC oxides \cite{kulic},
\cite{abrikosov}. However, for the quantitative theory the EPI-FSP
model must be refined by including the realistic phonon and band
structure of HTSC oxides.

Finally, based solely on the ARPES and tunnelling spectra, as well
as on dynamical conductivity measurements, one can make a reliable
\textit{phenomenological theory} for the pairing in HTSC oxides.
Its basic ingredient is the electron-phonon interaction (which
provides the strength for pairing) with the pronounced forward
scattering peak, whatsoever is its cause, while the Coulomb
scattering, in spite of its weakness compared to the EPI, triggers
superconductivity to d-wave like \cite{kulic}, \cite{kulic2},
\cite{kulic3}. This phenomenology is described by Eqs.(1-4), or
their generalization to a realistic forward scattering peak with
small but finite width $\delta \varphi _{w}\ll \pi $.

\textbf{Acknowledgment} - We thank Igor Mazin and O. Jepsen for careful
reading of the manuscript. M. L. K. thanks Ulrich Eckern, Peter Kopietz,
Igor and Lila Kuli\'{c} for support.

\end{document}